\documentclass[twocolumn,aps,prb,superscriptaddress]{revtex4}
\usepackage[latin1]{inputenc}
\usepackage{graphicx}

\usepackage{color}
\usepackage{bm}

\begin{document}

\title{On the nature of tunable hole g-factors in quantum dots}

\author{N.~Ares}
\affiliation{SPSMS, CEA-INAC/UJF-Grenoble 1, 17 Rue des Martyrs, F-38054 Grenoble Cedex 9, France}

\author{V.~N.~Golovach}
\affiliation{SPSMS, CEA-INAC/UJF-Grenoble 1, 17 Rue des Martyrs, F-38054 Grenoble Cedex 9, France}
\affiliation{Institute for Integrative Nanosciences, IFW Dresden,
 Helmholtzstr.\ 20, D-01069 Dresden, Germany}
 
\author{G.~Katsaros}
\affiliation{SPSMS, CEA-INAC/UJF-Grenoble 1, 17 Rue des Martyrs, F-38054 Grenoble Cedex 9, France}
\affiliation{Institute for Integrative Nanosciences, IFW Dresden,
 Helmholtzstr.\ 20, D-01069 Dresden, Germany}

\author{M.~Stoffel}
\affiliation{Universit\'e de Lorraine, Institut Jean Lamour, UMR CNRS 7198, Nancy-Universit\'e, BP 239, F-54506 Vandoeuvre-les-Nancy, France}

\author{F.~Fournel}
\affiliation{CEA, LETI, MINATEC, 17 Rue des Martyrs, F-38054 Grenoble Cedex 9, France}

\author{L.~I.~Glazman}
\affiliation{Department of Physics, Yale University, New Haven, Connecticut 06520, USA}

\author{O.~G.~Schmidt}
\affiliation{Institute for Integrative Nanosciences, IFW Dresden,
 Helmholtzstr.\ 20, D-01069 Dresden, Germany}

\author{S.~De~Franceschi}
\affiliation{SPSMS, CEA-INAC/UJF-Grenoble 1, 17 Rue des Martyrs, F-38054 Grenoble Cedex 9, France}

\begin{abstract}
Electrically tunable g-factors in quantum dots are highly desirable 
for applications in quantum computing and spintronics.
We report giant modulation of the hole g-factor in a SiGe nanocrystal
when an electric field is applied to the nanocrystal along its growth direction.
We derive a contribution to the g-factor that stems from an orbital effect of the magnetic field, 
which lifts the Kramers degeneracy in the nanocrystal by altering the mixing between the heavy and the light holes.
We show that the relative displacement between the heavy- and light-hole wave functions,
occurring upon application of the electric field,
has an effect on the mixing strength and
leads to a strong non-monotonic modulation of the g-factor.
Despite intensive studies of the g-factor since the late 50's,
this mechanism of g-factor control 
has been largely overlooked in the literature.
\end{abstract}

\date{\today{}}

\maketitle
In the past decade,
a great effort has been devoted to the realization of
spin qubits in semiconductors~\cite{Loss,Hanson}. 
Spin manipulation was achieved through different approaches:
magnetic-field-driven electron spin resonance~\cite{Koppens}, 
electric-dipole spin resonance~\cite{Nowack,Stevan,Golovach}, 
and fast control of the exchange coupling~\cite{Petta}. 
Another possibility for electric-field spin manipulation is the g-tensor modulation resonance, 
which has been used on ensembles of spins in two-dimensional (2D) electron systems~\cite{Kato,Salis}. 
This technique relies on anisotropic and electrically tunable g-factors. 
Recently, several experiments have addressed the g-factor modulation by means of external
electric fields~\cite{tarucha1,Finley1}, and different mechanisms were evoked to explain the observed g-factor tunability, such as 
compositional gradients~\cite{tarucha1} and quenching of the angular momentum~\cite{Finley1,PryorFlatte}.
Here we report the experimental observation of an exceptionally large and non-monotonic electric-field modulation of the hole g-factor in SiGe QDs. 
To interpret this finding we have to invoke a new mechanism that applies to hole-type low-dimensional systems. This mechanism 
relies on the existence of an important, yet overlooked correction term in the g-factor whose magnitude depends on the mixing of heavy and light holes. 
We show that in SiGe self-assembled QDs an electric field applied along the growth axis can be used to efficiently alter this mixing and produce 
large variations in the hole g-factor.

\begin{figure}
\includegraphics[width=1.0\columnwidth]{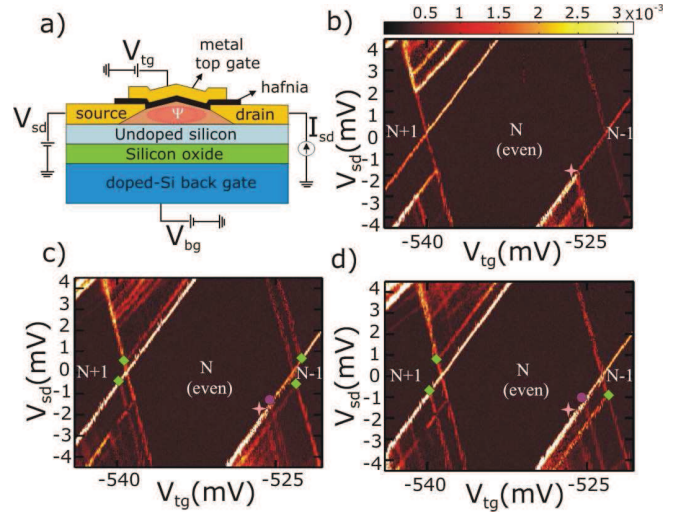}
\caption{\label{Fig1}
(a) Schematic cross section of a SiGe QD device. 
(b)-(d) Color plot of $dI_{\it sd}/dV_{\it sd}(V_{\it tg},V_{\it sd})$ for $B_{z}= 70\,{\rm mT}$, $3\,{\rm T}$ and $5\,{\rm T}$, respectively ($V_{\it bg} = 0$). 
The lines indicated by rhombis correspond to the onset of tunneling via Zeeman-split levels for $N-1$ and $N+1$ holes on the QD. 
The lines indicated by a star and by a circle correspond to singlet-triplet excitations for $N$ holes.
}
\end{figure}

Our SiGe QDs were grown by molecular-beam
epitaxy on a silicon-on-insulator substrate. 
The Stranski-Krastanow growth mode was tuned to yield dome-shaped QDs with height $w=20\,{\rm nm}$
and base diameter $d=80\,{\rm nm}$.  
A sketch of the device is shown in Fig.~\ref{Fig1}~(a).
The QD is contacted by two $20$-nm-thick Al electrodes, acting as source and drain leads.
A ${\rm Cr}/{\rm Au}$ gate electrode is fabricated on top of the QD with a 6-nm-thick hafnia interlayer 
deposited by atomic-layer deposition. This top gate, together with the degenerately-doped Si back gate, allows  
a perpendicular electric field to be applied while maintaining  a constant number of holes in the SiGe QD.

Measurements of the g-factor were performed using single-hole tunneling spectroscopy. A typical differential conductance
($dI_{\it sd}/dV_{\it sd}$) measurement as a function of top-gate voltage ($V_{tg}$) and
source-drain bias voltage ($V_{\it sd}$) is shown in Fig.~\ref{Fig1}~(b). All
measurements reported here were done in a ${}^{3}{\rm He}$ refrigerator with a base
temperature of $250\,{\rm mK}$. In order to suppress the superconductivity of the leads, 
a small magnetic field, $B_{z}=70\,{\rm mT}$, was applied along the $z$ axis, i.e. perpendicular to the $(x,y)$ growth plane.
Diamond-shaped regions, 
where the current vanishes due to Coulomb blockade, 
can be clearly observed in Fig.~\ref{Fig1}~(b).
The charging energy is about $10\,{\rm meV}$. 
Outside the diamonds,
additional lines denoting transport through excited orbital states 
can be observed. 
Figs.~\ref{Fig1}~(c) and (d) show the same Coulomb-blockade regime for 
$B_{z}=3\,{\rm T}$ and $B_{z}=5\,{\rm T}$, respectively. 
The magnetic field causes a splitting of the diamond edges as indicated by green rhombis. This splitting follows from the lifting of Kramers degeneracy in the ground states associated with the side diamonds. We thus conclude that the central diamond corresponds to an even number, $N$, of confined holes~\cite{Hanson}. The Zeeman energy splitting 
is given by $E_{Z}=g_{\perp}\mu_{B}B_{z}$, where $\mu_{B}$ is the Bohr magneton and $g_{\perp}$ is the
absolute value of the g-factor along $z$. 
From the splitting of $N$-hole diamond edges we extract $g_\perp=(3.0 \pm 0.4)$ and $g_\perp=(2.8 \pm 0.4)$ for the $N-1$ and the $N+1$ ground states, respectively. 
The line indicated by a star in Fig.~\ref{Fig1}~(b) is due to the spin-triplet excited state for $N$ holes on the QD. We measure a $2\,{\rm meV}$ singlet-triplet energy in this particular QD, which is an order of magnitude larger than for electrons in Si/SiGe heterostructures~\cite{Shaji}. 
We note that large singlet-triplet excitation energies are particularly desirable for the observation of spin blockade in double-dot experiments~\cite{OnoScience}.
Upon increasing $B_z$, the line denoted by a star splits as shown by the emergence of second parallel line, denoted by a circle, that shifts away proportionally to $B_z$ (see Figs.~\ref{Fig1}~(c) and (d)). This behavior corresponds to the Zeeman splitting of the excited spin-triplet state~\cite{Hanson} with $g_{\perp}=(2.8\pm 0.4)$. Hereafter, we will concentrate on g-factor measurements in spin-$1/2$ ground states.

\begin{figure}
\includegraphics[width=1.0\columnwidth]{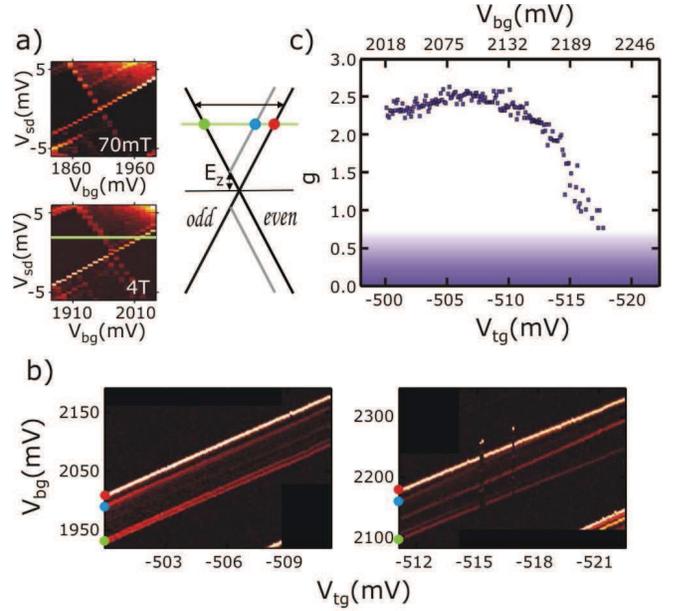}
\caption{\label{Fig2}
(a) Left: Color plots of $dI_{\it sd}/dV_{\it sd}(V_{\it bg},V_{\it sd})$ for $B=70\,{\rm mT}$ and $4\,{\rm T}$. 
At $4\,{\rm T}$ the Zeeman splitting is clearly visible. 
Right: Corresponding schematic diagram illustrating the measurement principle to extract the Zeeman energy splitting 
(and hence the g-factor) from gate-voltage sweeps at constant $V_{\it sd}$ (see the horizontal green line).
(b) Color plots of $dI_{\it sd}/dV_{\it sd}(V_{\it bg},V_{\it tg})$ for a fixed $V_{\it sd} = 2.6\,{\rm mV}$. 
These data sets demonstrate the modulation of $g_\perp$ 
by a perpendicular electric field proportional to $V_{\it bg} - V_{\it tg}$. 
(c) $g_\perp(V_{\it bg},V_{\it tg})$ as extracted from (b).
Below $g_\perp \approx 0.75$ the Zeeman splitting cannot be resolved any more due to the finite broadening of the tunneling resonances.
}
\end{figure}

Our dual-gate devices allow us to measure the dependence of the g-factor 
on a perpendicular electric field, $F$, at constant number of holes. The principle of such a
measurement is illustrated in Fig.~\ref{Fig2}~(a).
The Zeeman splitting is given by the distance between the blue and the red circles along $V_{\it bg}$, 
multiplied by a calibration factor $\alpha$. The latter is obtained by dividing $V_{\it sd}$ by the distance between the green and the red circles.
In order to investigate the $F$-dependence of the g-factor,
a constant $V_{\it sd} = 2.6\,{\rm mV}$ was
applied and $V_{\it bg}$ was swept while stepping $V_{\it tg}$. 
The magnetic field was fixed at $4\,{\rm T}$.  
The data is shown in Fig.~\ref{Fig2}~(b) and the extracted
g-factors are displayed in Fig.~\ref{Fig2}~(c). 
We observe an exceptionally large g-factor modulation ($\delta g/g \sim 1$) 
denoting a strong effect of the applied $F$. 
The g-factor increases slowly to a maximum value of $2.6$ and then drops rapidly down to a point where the Zeeman
splitting can no longer be resolved. 
Comparably large g-factor variations have been observed in other similar measurements, see Appendix~\ref{SupplExp}.

In order to uncover the origin of this unusual behavior, we modelled the QD electronic states in terms of heavy-hole (HH) and light-hole (LH) subbands. 
Given the relatively large anisotropy of dome-shaped QDs, we initially considered the two-dimensional (2D) limit resulting from confinement along the growth axis. 
Due to quantum confinement along $z\equiv [001]$ and strain, the 4-fold degeneracy of the valence band at $\Gamma$-point is lifted.
The top-most subband has HH character 
and its in-plane dispersion relation is described by the effective 2D Hamiltonian
\begin{eqnarray}
H_{\rm eff} &=& \frac{1}{2m_\parallel}\left(k_x^2+k_y^2\right) + 
\frac{1}{2}g_\parallel\mu_B\left(\sigma_xB_x+\sigma_yB_y\right)\nonumber\\
&&-\frac{1}{2}g_\perp\mu_B\sigma_zB_z+U(x,y),
\label{Heff2Dhh}
\end{eqnarray}
where $k_x$ and $k_y$ are the in-plane momentum operators, 
$m_\parallel=m/(\gamma_1+\gamma_2)$ is the in-plane effective mass~\cite{Winkler},
$g_\parallel = 3q$ and $g_\perp=6\kappa+\frac{27}{2}q$ are, respectively, 
the in-plane and transverse g-factors~\cite{Winkler,vanKesteren},
$\bm{\sigma}$ are the Pauli matrices in the pseudospin space~\cite{note1},
and $U(x,y)$ is the in-plane confining potential in the QD.
We use standard notations for the Luttinger parameters 
$\gamma_1$, $\gamma_2$, $\gamma_3$, $\kappa$, and $q$~\cite{Luttinger}.
Since $q\ll \kappa$, it is appropriate to assume $g_\perp\approx 6\kappa$.
The minus sign in front of $\frac{1}{2}g_\perp$ in Eq.~(\ref{Heff2Dhh})
is introduced for the convenience of having $g_\perp$ positive for Ge.

First we consider the possibility that the observed g-factor modulation arises from a compositional gradient.  

It is well known that Si and Ge intermix leading to the formation of a ${\rm Si}_{1-{\sf x}}{\rm Ge}_{\sf x}$ alloy
in which ${\sf x}$ increases motonically with $z$, being zero at the base ($z=-w$) and approaching unity at the apex ($z=0$) of the QD~\cite{Schulli}.
Since $\kappa_{\rm Si}=-0.42$ and $\kappa_{\rm Ge}=3.41$, one would expect that $g_{\perp}$ increases with $F$ 
following a vertical shift of the HH wave function towards the apex. 
This {\em compositional-gradient} mechanism was exploited 
in ${\rm Al}_{\sf x}{\rm Ga}_{1-{\sf x}}{\rm As}$ 
quantum wells to implement electrical control of electron spins~\cite{Salis,Kato}.
While for electrons this may well be the only efficient way to control the g-factor,
the situation might be different for holes.

\begin{figure}
\includegraphics[width=1.0\columnwidth]{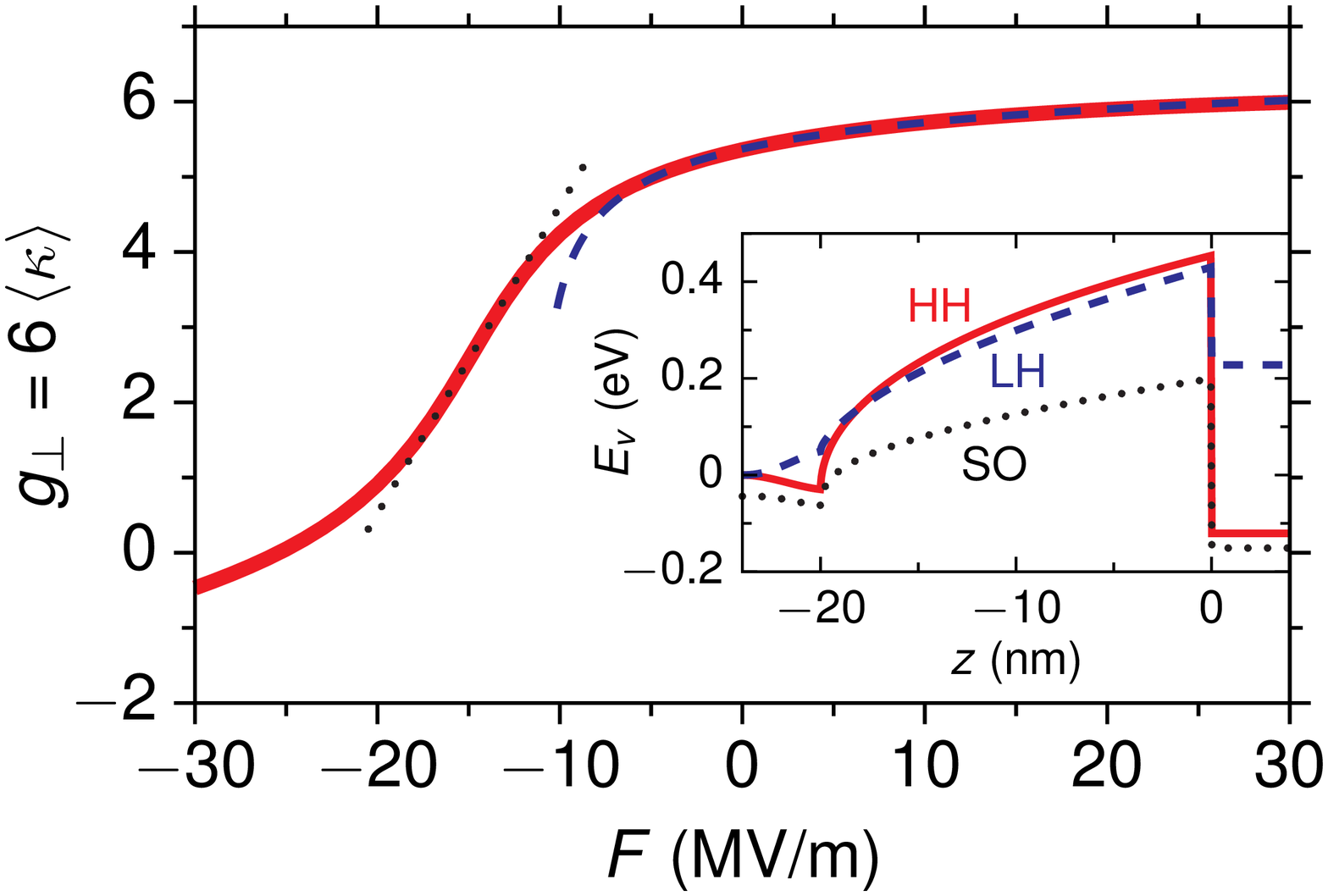}
\caption{\label{Fig3}
Expected electric-field dependence of $g_\perp$ 
for a SiGe QD with a strong compositional gradient.
The numerical result (solid line) has two regimes, designated by fits
to two simplistic models.
The dashed line shows a fit to the expression
$\left\langle\kappa\right\rangle=\kappa_\infty -\Delta\kappa\left(1+F/F_{\rm intr}\right)^{-1/3}$,
derived for a triangular potential well with an intrinsic electric field $F_{\rm intr}$ for $z<0$
and infinite barrier for $z>0$.
The dotted line shows a fit to a linear dependence, obtained for a symmetric potential well.
In the latter regime, $g_\perp$ is most sensitive to $F$, with
$dg_\perp/dF=0.41\,{\rm m}/{\rm MV}$.
At large negative $F$, the wave function is pushed into the Si-rich region, where $g_\perp$ becomes negative.
(Inset) Energy profiles for the heavy-hole (HH), light-hole (LH), and split-off (SO) bands for a Ge content ${\sf x}$ given by Eq.~(\ref{eqnxcompsqrt})
with ${\sf x}_{\rm max}=0.8$. We complemented the QD model with a thin layer of
Si substrate at $z\in\left[-24,-20\right]\,{\rm nm}$
and a strained Si capping layer at $z\in\left[0,4\right]\,{\rm nm}$.
Strain is taken into account resulting in a splitting between HH and LH bands.
}
\label{fig6kap}
\end{figure}

To find an upper bound for the g-factor variation resulting from the 
compositional gradient, we take the steepest dependence reported for the Ge content across the QD~\cite{Schulli},
\begin{equation}
{\sf x}(z)={\sf x}_{\rm max}\sqrt{1+\frac{z}{w}},
\quad\quad\quad
-w < z < 0,
\label{eqnxcompsqrt}
\end{equation}
where $w=20\,{\rm nm}$ is the height of the dome-shaped QD.
To account for the existing uniaxial strain,
we assume that the in-plane lattice constant $a_\parallel$
increases linearly 
from $5.47\,\mbox{\rm \AA}$ at the base to
$5.59\,\mbox{\rm \AA}$ at the apex~\cite{Schulli}.
With these two ingredients, the valence band profiles 
$E_v(z)$ for all types of holes are calculated using
interpolation schemes devised for SiGe~\cite{WalleMartin,RiegerVogl}
(see inset of Fig.~\ref{Fig3}).
The HH ground state is thus confined to a triangular potential well arising
from the compositional gradient. An electric 
field applied along $z$ adds a term $-eFz$ to $E_v(z)$.
For a given $F$, the HH wave function $\psi(z)$ is obtained by
solving the Schr\"{o}dinger equation numerically.
The HH g-factor is found as a weighted average 
\begin{equation}
g_\perp\approx 6\left\langle \kappa \right\rangle = 6 \int \kappa\left[{\sf x}(z)\right]\left|\psi(z)\right|^2dz,
\label{eqn6kapint1}
\end{equation}
where $\kappa\left({\sf x}\right)$ is obtained using the non-linear interpolation described in Ref.~\onlinecite{WinklerCRGeSi}.
The resulting dependence $g_\perp (F)$ is shown in Fig.~\ref{Fig3}.
We distinguish two regimes: that of a strongly asymmetric (triangular) potential well
and that of a symmetric potential well.
The modulation of the g-factor is largest in the latter regime (see dotted line in Fig.~\ref{Fig3});
we obtain $d g_\perp/dF\approx 0.41\,{\rm m}/{\rm MV}$.
While the magnitude of the modulation is close to what is observed in the experiment,
the sign of $d g_\perp/dF$ is nevertheless opposite (see Fig.~\ref{Fig2}~(c)).
We conclude that the compositional-gradient cannot explain our data.
Therefore, from now on, we shall discard this mechanism and assume the Ge content to be constant within the QD.

We revisit the derivation of Eq.~(\ref{Heff2Dhh}), starting from the Luttinger Hamiltonian.
In the 2D limit, the $4\times4$ Luttinger Hamiltonian separates into $2\times2$ HH and LH blocks, see Appendix~\ref{SupplThr}.
To leading order of $w/d\ll 1$,
the HH and LH sectors are connected by the off-diagonal mixing blocks~\cite{spinselective}
\begin{equation}
H_{\it hl}=\left(H_{\it lh}\right)^\dagger =
i\frac{\sqrt{3}\gamma_3}{m}
\left( k_x\sigma_y + k_y\sigma_x\right)k_z,
\label{Hlh0order}
\end{equation}
where $k_x$ and $k_y$ are 2D versions of momentum operators 
(insensitive to in-plane magnetic fields),
$k_z \equiv -i\hbar\partial/\partial z$, and
$\sigma_x$ and $\sigma_y$ are
the Pauli matrices in a pseudospin space, introduced simultaneously for
heavy and light holes~\cite{spinselective}.

The mixing blocks in Eq.~(\ref{Hlh0order}) are proportional to $k_z$.
In spite of the fact that $k_z$ averages to zero for each type of hole separately,
it cannot be discarded in Eq.~(\ref{Hlh0order}),
because matrix elements of the type
$\left\langle \psi_h\right|k_z\left|\psi_l\right\rangle$
are, in general, non-zero and scale as $1/w$ for $w\to 0$.
Here, $\psi_h(z)$ and $\psi_l(z)$
obey two separate Schr\"{o}dinger equations, for heavy and light holes, respectively (see below).
This observation allows us to anticipate that in second-order perturbation theory
the mixing blocks lead to an energy correction containing 
$H_{\it hl}H_{\it lh}\propto k_z^2$ in the numerator and
$H_{\it ll}-H_{\it hh} \propto k_z^2$ in the denominator.
This correction does not vanish in the 2D limit ($k_z\to\infty$).
At the same time, the correction to the wave function vanishes as $k_\parallel/k_z\sim w/d$.

Using second-order perturbation theory, we recover Eq.~(\ref{Heff2Dhh}) for the top-most hole subband.
Yet, at the leading (zeroth) order of $w/d\ll 1$,
we obtain the following  modified expressions for the effective mass and perpendicular g-factor,
\begin{equation}
m_\parallel=\frac{m}{\gamma_1+\gamma_2-\gamma_h},
\quad\quad
g_\perp = 6\kappa+\frac{27}{2}q - 2\gamma_h.
\label{eqnmgzzgamh}
\end{equation}
The in-plane g-factor remains  unchanged ($g_\parallel=3q$) at this order.
In Eq.~(\ref{eqnmgzzgamh}), $\gamma_h$ is a dimensionless parameter sensitive to the form of the confinement along $z$,
\begin{equation}
\gamma_h = \frac{6\gamma_3^2}{m}\sum_{n}
\frac{\left|\left\langle \psi_n^l\right|k_z\left|\psi_1^h\right\rangle\right|^2}{E_n^l-E_1^h}.
\label{gamheqdef2}
\end{equation}
Here, the sum runs over the LH subbands and
the wave functions $\psi_n^{h/l}(z)$ and energies $E_n^{h/l}$
obey 
\begin{equation}
\left[\frac{k_z^2}{2m_\perp^{h/l}}+V_{h/l}(z)\right]\psi_n^{h/l}(z) = E_n^{h/l}\psi_n^{h/l}(z),
\label{1schrEqshl}
\end{equation}
where $m_\perp^{h/l}=m/(\gamma_1\mp 2\gamma_2)$ and 
$V_{h/l}(z)$ is the confining potential seen by the heavy/light hole.

When $V_h(z)$ and $V_l(z)$ are infinite square wells, 
the quantity $\gamma_h$ in Eq.~(\ref{gamheqdef2}) can be derived analytically,
\begin{equation}
\gamma_h = \frac{12\gamma_3^2}{\gamma_1+2\gamma_2}
\left[\frac{1}{1-\beta}
-\frac{4\sqrt{\beta}}{\pi\left(1-\beta\right)^2}\cot\left(\frac{\pi}{2}\sqrt{\beta}\right)\right],
\label{eqnhammahsquarewellpot}
\end{equation}
where $\beta = m_\perp^l/m_\perp^h + \delta E_{001}/E_1^l$, 
with $\delta E_{001}\equiv V_h - V_l$ 
being the splitting of the valence band due to uniaxial strain
and $E_1^l=\pi^2\hbar^2/2m_\perp^lw^2$.
Notably, one has $\psi_n^h(z) = \psi_n^l(z)$ in this case, because the masses
$m_\perp^h$ and $m_\perp^l$ drop out of the expressions for the wave functions.
With the application of an electric field, 
the wave functions $\psi_n^h$ and $\psi_n^l$ begin to shift relative to each other, 
because of their different effective masses.
Although $\gamma_h$ can only be numerically computed,
its qualitative dependence on $F$ can be inferred by inspecting Eq.~(\ref{gamheqdef2}).
Note that the $n=1$ term in the sum has the smallest energy denominator 
and, therefore, it is expected to have a dominant contribution.
For a square-well potential, however, this term vanishes by symmetry.
As a result, the symmetric point $F=0$ corresponds to a minimum in $\gamma_h(F)$,
since $E_n^l>E_1^h$.
Away from $F=0$, $\gamma_h$ increases quadratically, $\gamma_h\propto F^2$, 
up to the point where the electric field is strong enough to shift the HH wave function ($eFw\simeq E_2^h-E_1^h$).
Then, $\gamma_h$ increases roughly linearly up to the point where the LH wave functions begin to shift ($eFw\simeq E_2^l-E_1^l$).
Upon further increasing $F$, $\gamma_h$ increases weakly and saturates to a constant.
We remark that $g_\perp$ is modified by $\gamma_h$
even at $k_\parallel=0$, despite the fact that no HH-LH mixing occurs at $k_\parallel=0$.
In fact, $g_\perp$ is sensitive to orbital motion in a perpendicular magnetic field~\cite{Matveev}, and even a small $B_z$ translates to $k_\parallel\neq 0$, 
leading to HH-LH mixing.

\begin{figure}
\includegraphics[width=1.0\columnwidth]{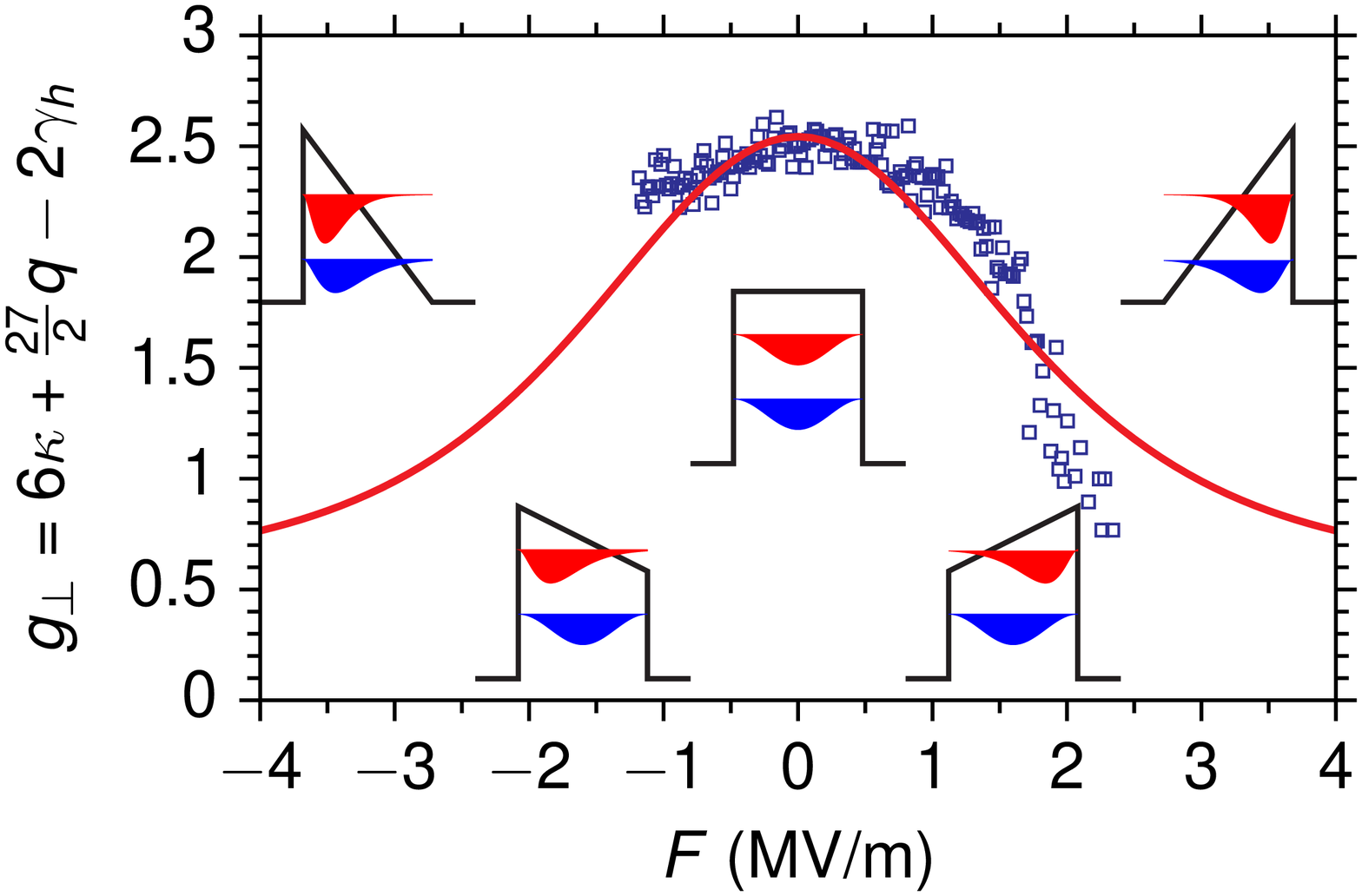}
\caption{
The $g_\perp$($F$) dependence according to Eq.~(\ref{eqnmgzzgamh}) (red line) and corresponding 
experimental data (open dots).
(Insets) Schematics of the HH and LH confinement at different $F$.
The central inset shows the symmetric-well configuration at $F=0$.
The upper insets show the regime of strong $F$ where both types of holes are confined to one side of the (triangular) potential well.
The lower insets show the intermediate regime in which $F$ is strong enough to
push the heavy hole to one side of the well, 
while the light hole remains spread over the entire well width.
}
\label{figgzzvsF}
\end{figure}

Our result in Eq.~(\ref{eqnmgzzgamh}) represents the zeroth-order term in the
expansion $g = g^{(0)}+ g^{(2)} + \dots$, where $g^{(2)} \propto (w/d)^2$ is the subleading-order term.
Unlike the main term,
the correction $g^{(2)}$ is sensitive to the in-plane confining potential $U(x,y)$
and it originates from the HH-LH interference terms in the wave function.
We shall address $g^{(2)}$ in a separate work.
In Fig.~\ref{figgzzvsF}, we fit the experimental data
using only the leading, zero-order term. 
The insets in Fig.~\ref{figgzzvsF} illustrate how the HH and LH wave functions, $\psi_1^h(z)$ (red) and $\psi_1^l(z)$ (blue),
shift upon application of the electric field.
Note that the transition from the square well (central inset) to the triangular well (upper insets) occurs in two steps.
First, the HH wave function shifts by $\delta z\sim w$,
while the LH wave function remains nearly unaffected (lowest insets).
Then, the LH wave function shifts as well (highest insets).
At even larger $F$ (not shown) $g_\perp$ saturates to $g_\perp\approx 0.6$.
The evolution of $g_\perp$ taking into account $\gamma_h$, shown as a solid red line in Fig.~\ref{figgzzvsF}, qualitatively reproduces our experimental results (open dots in Fig.~\ref{figgzzvsF}).

Finally, we remark that the correct 2D limit of the Luttinger Hamiltonian has been largely overlooked in the literature on 2D hole systems.
Nevertheless, our main result in Eq.~(\ref{eqnmgzzgamh}) bears some relation to earlier works.
D'yakonov and Khaetskii~\cite{DyakonovKhaetskii} studied the Luttinger Hamiltonian in an infinite square well 
and used the spherical approximation ($\gamma_2=\gamma_3$).
They derived an expression for $m_\parallel$ that agrees with our result in the appropriate limit.
We also verified that Eqs.~(\ref{eqnmgzzgamh}) and (\ref{gamheqdef2}) can be obtained from a general $\bm{k\cdot{p}}$-approach~\cite{Roth,KiselevIvchenkoRossler} after a lengthy calculation.
In spite of the previous work, however,
the relation of $m_\parallel$ and $g_\perp$ to an additional parameter $\gamma_h$
and the fact that $\gamma_h$ is sensitive to $F$
have been missing from the general knowledge of 2D hole systems.

In conclusion, we showed that an external electric field can strongly
modulate the perpendicular hole g-factor in SiGe QDs. 
By a detailed analysis, we ruled out the compositional-gradient mechanism as the origin of this 
electric-field effect. 
By analyzing the Luttinger Hamiltonian in the 2D limit, we
found a new correction term $\gamma_h$ which had not been considered before in
the literature. 
This new term, which corrects the ``standard'' expression for the HH
g-factor, reflects the effect of a perpendicular magnetic-field on the orbital motion, and it is ultimately related to the atomistic spin-orbit coupling
of the valence band. 

We acknowledge financial support from
the Nanosciences Foundation (Grenoble, France), DOE under the Contract No.~DEFG02-08ER46482 (Yale), the Agence Nationale de la Recherche, and the European Starting Grant. 

\appendix

\section{Supplementary data on the g-factor modulation}
\label{SupplExp}

\begin{figure}
\includegraphics[width=1.0\columnwidth]{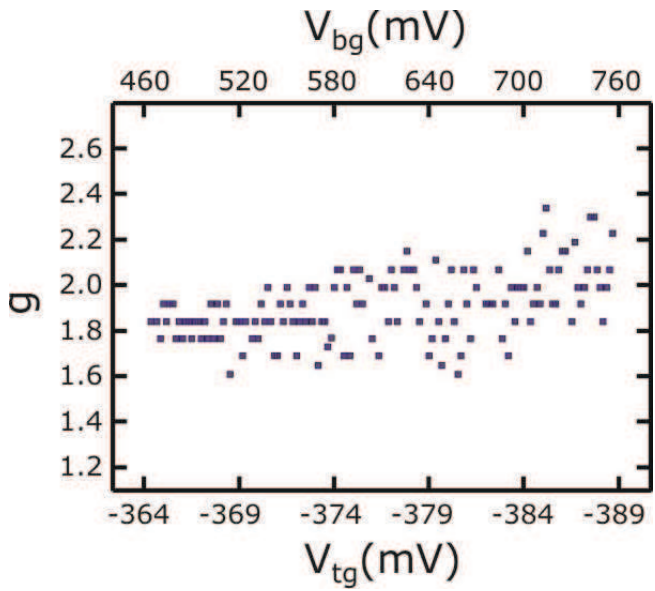}
\caption{
Measured g-factor versus $V_{\it tg}$ and $V_{\it bg}$ for the triangular potential well configuration. 
The dispersion of the data points is too big to resolve a clear variation in the g-factor value. 
\label{SFig1}}
\end{figure}

According to the given theoretical description of the g-factor variation, we should be able to observe a strong effect  
just when the applied electric field drives the system close to the flat-band configuration. 
This is the case exhibited in Fig.~\ref{figgzzvsF} of the main text. 
In Fig.~\ref{SFig1} the g-factor variation is shown for the same device but in an electric field regime in which the potential well configuration is triangular. 
Although the range of $V_{\it tg}$ and $V_{\it bg}$ is similar in both Fig.~\ref{figgzzvsF} and Fig.~\ref{SFig1}, 
there is no obvious change in the g-factor value for the latter case, as expected.
  
It is known that thermal cycling can change the characteristics of a device and
thus different set of data can be obtained for the same device~\cite{tarucha}.
A drastic change in $V_{\it tg}$ (of about $3\,{\rm V}$) appeared to produce a similar
effect in our system. In this way, we could perform the g-factor measurements
on two devices with completely different characteristics. This time, a slightly different type of measurement, 
which will be explained below, was performed in order to extract the g-factors.

\begin{figure}[t]
\includegraphics[width=8cm]{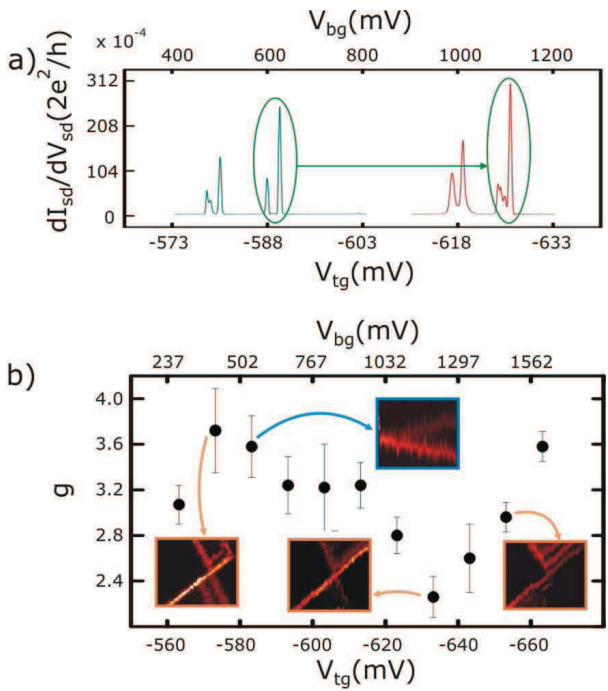}
\caption{
(a) Plot of
$dI_{\it sd}/dV_{\it sd}$ versus $V_{\it bg}$ for  $V_{\it sd}= 1\,{\rm mV}$ 
showing how two coulomb peaks move while changing the value of the external electric field. 
(b) Plot of the g-factor vs $V_{\it bg}$ and $V_{\it tg}$ showing a non monotonic behavior of the
g-factor value for different perpendicularly applied electric fields. 
The insets show stability diagrams at $B=1.5\,{\rm T}$. 
$V_{\it bg}$ is swept by $50\,{\rm mV}$ while $V_{\it sd}$ by $2.4\,{\rm mV}$. 
The blue inset demonstrates that indeed the parallel to the ground state line is 
due to the Zeeman splitting, because it merges with the ground state for vanishing magnetic fields. 
\label{SFig2}}
\end{figure}

As discussed in the main text, by having a dual gate configuration it is
possible to keep the same number of holes while changing the value of the
external electric field. 
The same Coulomb peak can be very easily followed
since for each ${\rm mV}$ we move in $V_{\it tg}$, the peaks move $\sim 13.2\,{\rm mV}$ in
$V_{\it bg}$. 
This ratio reflects the difference in coupling between the QD and the
two gates.  
Fig.~\ref{SFig2}~(a) shows how two coulomb peaks shift in $V_{\it bg}$
while changing $V_{\it tg}$ from approximately $-583\,{\rm mV}$ (blue trace) to
$V_{\it tg}\simeq -620\,{\rm mV}$ (red trace). 
The reason why the Coulomb peaks appear to
be split is that $dI_{\it sd}/dV_{\it sd}$ was measured under 
$V_{\it sd}= 1\,{\rm mV}$.

By following particular Coulomb peaks we could, for different values of
$V_{\it tg}$, sweep $V_{\it sd}$ as a function of $V_{\it bg}$, 
and obtain a stability diagram
for the same hole state under different applied electric fields.  From the
Zeeman-split lines observed in the stability diagrams, we have obtained the
g-factors for $11$ different values of $V_{\it tg}$ and $V_{\it bg}$, see
Fig.~\ref{SFig2}~(b). The inset which is surrounded by a blue rectangle shows
the $dI_{\it sd}/dV_{\it sd}$ as a function of $(B, V_{\it sd})$ for 
$V_{\it tg}\simeq -583.2\,{\rm mV}$. 
This plot demonstrates that the parallel to the ground state line seen in
the stability diagrams insets (orange rectangles), is indeed the Zeeman
splitting. 
From Fig.~\ref{SFig2}~(b) it becomes clear that the external electric field, 
has a very strong effect on the g-factor value.  
It reaches a minimum value of $(2.3 \pm 0.2)$ and a maximum value of
$(3.7 \pm 0.4)$. This is, to our knowledge, the biggest g-factor value
measured in SiGe nanostructures. Even larger g-factors can be obtained for QDs
with higher Ge content, as has been shown by theoretical calculations~\cite{Nenashev}.

Interestingly, two different behaviours are present in this set of data. From
$V_{\it tg}\approx-560$ to $V_{\it tg}\approx-630$, the modulation in the g-factor
resembles the one shown in the main text.
It can be explained by the dependence of $\gamma_h$ on $F$.
However, from $V_{\it tg}\approx-630$ to $V_{\it tg}\approx-660$
an abrupt increase of the g-factor for increasing electric field can be observed. 
This behaviour is not compatible with the explanation given in the main text.
A careful inspection of the stability diagram of Fig.~\ref{SFig2}~(b) for
$V_{\it tg}\approx -650$ shows that additional levels are present close
the studied level.
No closely-lying levels are present in the range $V_{\it tg}\in\left[-560,-630\right]$.
We believe that the data in the range $V_{\it tg}\in\left[-630,-660\right]$
refers to a different size-quantization level than the data in the range $V_{\it tg}\in\left[-560,-630\right]$.
Indeed, the ground-state level might have changed in going from one range to the other,
which is presumably signified by the pronounced cusp in the g-factor data 
at $V_{\it tg}\approx -630\,{\rm mV}$ in Fig.~\ref{SFig2}~(b).
Such transitions are expected to occur when the Fermi level is placed sufficiently
deep in the valence band.
Then, ladders of levels belonging to different heavy-hole subbands move with respect to each other
when the electric field is varied.
Since these levels have very different $z$-components of the wave function, they interact weakly and
can come close to each other without a sizable level repulsion.
Levels belonging to one ladder cannot come close to each other when the electric field is varied.
We therefore speculate that the data in the range $V_{\it tg}\in\left[-630,-660\right]$
refers to a level from the second heavy-hole subband.

The following question arises: 
How can the two heavy-hole subbands have different characteristic values of the electric field at which the g-factor reaches maximum.
It is important to note that our model in the main text is strongly simplified.
The real confining potentials $V_h(z)$ and $V_l(z)$ in the NC are, most likely, never perfectly symmetric
and differ from each other.
Therefore, it is natural to expect that the alignment along $z$
of wave functions of the two heavy-hole subbands 
with respect to each other and, at the same time, with respect to the light-hole subbands
is not perfect. 
The data in Fig.~\ref{SFig2}~(b) is consistent with the
assumption that the wave function of the second heavy-hole subband
is shifted towards the base of the NC and it begins to align with the
light-hole wave functions at a later value of the electric field.

It is important to remark that higher-order corrections to the expansion 
of the g-factor of the top-most subband 
cannot explain the abrupt increase 
taking place for $V_{\it tg}\in\left[-630,-660\right]$,
because their contribution is small by $\sim (w/d)^{2}$, 
which amounts to only a $10\%$-correction to the main term for our devices.

\section{The 2D limit of the Luttinger Hamiltonian}
\label{SupplThr}

Here we derive the expression for $\gamma_h$ given in Eq.~(6) of the main text.
Our derivation is based on an expansion of the Luttinger Hamiltonian 
around the two-dimensional (2D) limit,
which we have recently outlined in Ref.~\onlinecite{spinselective}.
We summarize briefly the relevant results of Ref.~\onlinecite{spinselective}
and address the following question: 
What is the effective Hamiltonian of the top-most hole subband in the 2D limit?

We start with the representation of the Luttinger Hamiltonian in block form,
\begin{equation}
H=
\left(
\begin{array}{cc}
H_{\it hh} & H_{\it hl}\\
H_{\it lh} & H_{\it ll}
\end{array}
\right),
\label{HamblockHhhHllHhlHlh}
\end{equation}
where $H_{\it hh}$ and $H_{\it ll}$ are the main blocks, describing  heavy holes and light holes, respectively.
The off-diagonal blocks $H_{\it hl}$ and $H_{\it lh}\equiv \left(H_{\it hl}\right)^\dagger$ 
give the mixing between the heavy-hole and light-hole sectors and are responsible for
avoided crossings which occur between heavy-hole and light-hole branches at higher energies.
The energy axis is chosen to point downwards for holes. 
In the 2D limit, the heavy-hole and light-hole sectors become well separated in energy and 
the top-most hole subband (lowest in energy) is described by purely heavy-hole states,
i.e. states containing no admixture from the light-hole sector.
It is, therefore, customary~\cite{YuCardona} 
to neglect the off-diagonal blocks in Eq.~(\ref{HamblockHhhHllHhlHlh}) 
when considering the 2D limit and describe the top-most hole subband by
an effective Hamiltonian derived only from the block $H_{\it hh}$,
\begin{equation}
H_{\rm eff} = \left\langle H_{\it hh}\right\rangle,
\label{Heffwrong}
\end{equation}
where $\left\langle\dots\right\rangle$ stands for averaging over the motion along $z$.
Contrary to the common expectation, Eq.~(\ref{Heffwrong}) does not describe correctly the 2D limit of the Luttinger Hamiltonian.
The correct expression in the place of Eq.~(\ref{Heffwrong}) involves terms of the order of $H_{\it hl}^2/H_{\it ll}$, 
which remain finite in the 2D limit.
While these terms do not change the qualitative picture of the result, they are important when mechanisms of 
electric control of the g-factor and effective mass are considered.

Before proceeding further we need to clarify what exactly we mean by the 2D limit.
By the 2D limit of the Luttinger Hamiltonian, 
we mean a limit in which the $z$-sizes of both heavy-hole and light-hole wave functions
are sent to small values, e.g. by means of confining the holes to a thin layer of material.
Of course, the Luttinger Hamiltonian is by itself an effective Hamiltonian (intended for top of valence band)
and it is valid only for energies much smaller than the spin-orbital energy $\Delta_{\rm SO}$.
However, $\Delta_{\rm SO}$ is, typically, a large energy and one can envision the following limit,
\begin{equation}
E\ll\Delta_z\ll\Delta_{\rm SO},
\label{eqn2DlimEDzDSO}
\end{equation}
where $E$ is the energy measured away from the edge of the top-most subband 
and $\Delta_z\sim \left(\gamma_2/m\right)\left\langle k_z^2\right\rangle$ is the splitting energy between heavy and light holes due to 
confinement (or due to strain if strain dominates).
We, therefore, have in mind Eq.~(\ref{eqn2DlimEDzDSO}) when talking about the 2D limit of the Luttiger Hamiltonian.
It is important to note that, in the Luttinger Hamiltonian~\cite{Luttinger}, 
$\Delta_{\rm SO}$ is already sent to infinity and, thus, the second inequality in
Eq.~(\ref{eqn2DlimEDzDSO}) is, formally, fulfilled within the model.
In practice, however, our derivation remains qualitatively correct up to $\Delta_z\lesssim \Delta_{\rm SO}$.
For Ge, one has $\Delta_{\rm SO}\approx 0.3\,{\rm eV}$.

After expanding the Luttinger Hamiltonian in terms of the small parameter $w/d\ll 1$
(see Ref.~\onlinecite{spinselective}),
the main blocks in Eq.~(\ref{HamblockHhhHllHhlHlh}) 
are given at the leading order by
\begin{eqnarray}
H_{\it hh}&=&\frac{\gamma_1+\gamma_2}{2m}\left(k_x^2+k_y^2\right)+\frac{\gamma_1-2\gamma_2}{2m}k_z^2
\nonumber\\
&&+\frac{1}{2}\mu_B\bm{\sigma}\cdot g_h\cdot\bm{B}+U(x,y)+V_h(z),\nonumber\\
H_{\it ll}&=&\frac{\gamma_1-\gamma_2}{2m}\left(k_x^2+k_y^2\right)+\frac{\gamma_1+2\gamma_2}{2m}k_z^2
\nonumber\\
&&+\frac{1}{2}\mu_B\bm{\sigma}\cdot g_l\cdot\bm{B}+U(x,y)+V_l(z),
\label{eqnHhhHhl0oder}
\end{eqnarray}
where $k_x$ and $k_y$ are the 2D momentum operators, which contain only the $z$-component of the magnetic field.
The inplane components of the magnetic field contribute to $H$ 
at higher orders of the expansion, 
but not at the leading order considered here.
The g-factors, $g_h$ and $g_l$, entering Eq.~(\ref{eqnHhhHhl0oder}),
are diagonal in the frame $(x,y,z)$ and given by
\begin{equation}
g_h=
\left(
\begin{array}{ccc}
3q & 0 & 0 \\
0 & 3q &0 \\
0 & 0 & -6\kappa -\frac{27}{2}q
\end{array}
\right),
\end{equation}
and
\begin{equation}
g_l=
\left(
\begin{array}{ccc}
4\kappa+10q & 0 & 0 \\
0 & 4\kappa+10q &0 \\
0 & 0 & 2\kappa +\frac{1}{2}q
\end{array}
\right).
\end{equation}
The confining potential is assumed to separate into an inplane component $U(x,y)$ 
and a transverse $V_{h/l}(z)$.
The confining potential can be different for heavy and light holes, because of the strain.

The off-diagonal blocks are related to each other by hermiticity,
\begin{equation}
H_{hl}=\left(H_{lh}\right)^\dagger.
\end{equation}
For $H_{lh}$, we keep the leading-order term
\begin{equation}
H_{lh} =
-i\frac{\sqrt{3}\gamma_3}{m}
\left( k_x\sigma_y + k_y\sigma_x\right)k_z.
\label{SHlh0order}
\end{equation}
In writing Eq.~(\ref{eqnHhhHhl0oder}) and~(\ref{SHlh0order}), we made a choice of basis for the hole pseudo-spin.
We chose the basis such that the pseudo-spin transforms under the time-reversal operation as a spin $1/2$, 
see Ref.~\onlinecite{spinselective},
\begin{eqnarray}
\left|\uparrow\right\rangle_h&=&\left|3/2,-3/2\right\rangle, \quad
\left|\downarrow\right\rangle_h=\left|3/2,+3/2\right\rangle,\nonumber\\
\left|\uparrow\right\rangle_l&=&\left|3/2,+1/2\right\rangle, \quad
\left|\downarrow\right\rangle_l=\left|3/2,-1/2\right\rangle.
\label{basisps}
\end{eqnarray}
This choice of basis is convenient since it allows us to have 
``standard'' time-reversal transformations for all operators.
In Eq.~(\ref{basisps}), we use basis of states of the angular momentum $J=3/2$~\cite{Abakumov}:
\begin{eqnarray}
\left|\frac{3}{2},+\frac{3}{2}\right\rangle&=&-\frac{1}{\sqrt{2}}\left(X+iY\right)\uparrow,\nonumber\\
\left|\frac{3}{2},-\frac{3}{2}\right\rangle&=&\frac{1}{\sqrt{2}}\left(X-iY\right)\downarrow,\nonumber\\
\left|\frac{3}{2},+\frac{1}{2}\right\rangle&=&\frac{1}{\sqrt{6}}\left[-\left(X+iY\right)\downarrow+2Z\uparrow\right],\nonumber\\
\left|\frac{3}{2},-\frac{1}{2}\right\rangle&=&\frac{1}{\sqrt{6}}\left[\left(X-iY\right)\uparrow+2Z\downarrow\right],
\label{hhllbasis0}
\end{eqnarray}
where the functions $X$, $Y$, and $Z$ are real and represent the Bloch amplitudes of the valence band 
in the absence of spin-orbit interaction.

Next, we rotate away the off-diagonal blocks in Eq.~(\ref{HamblockHhhHllHhlHlh}) 
using perturbation theory~\cite{LandauLifshitz}.
We are allowed to do so if we focus on a heavy hole state which is
far away from any light hole states.
This requirement is fulfilled in an energy window between the edges of the first heavy-hole and light-hole subbands, i.e. close to the top of the valence band.
Of course, occasionally, such a premise may hold for a state deep in the valence band. 
And, in particular, it may hold for a light-hole state.
In the latter case, we would be deriving an effective Hamiltonian in the favour of the light hole.
However, the majority of states deep in the valence band do not permit application of perturbation theory.

In order to quantify the applicability of our method, we introduce characteristic values 
for the momenta operators in a localized state:
\begin{equation}
\bar{k}_\alpha^h = \sqrt{\left\langle h\right|k_\alpha^2\left|h\right\rangle},\quad\quad (\alpha=x,y,z),
\end{equation}
where $\left|h\right\rangle$ stands for the localized (heavy-hole) state.
Then, the expansion around the 2D limit (i.e. separation of variables)
is valid if
\begin{equation}
\bar{k}_{x,y}^h \ll \bar{k}_z^h,
\end{equation}
whereas the perturbation theory can be applied if
\begin{equation}
\frac{\gamma_3}{m}\sqrt{\bar{k}_{x,y}^h \bar{k}_{x,y}^l} \ll E_l-E_h,\quad\quad \forall l,
\end{equation}
where the index $l$ refers to the light-hole states.
Here, $E_h$ and $E_l$ are energies of the heavy-hole and light-hole states, respectively.
Since we restrict our consideration to small inplane momenta,
we can approximate $E_h$ and $E_l$ by
\begin{eqnarray}
E_h &\approx& \frac{\gamma_1-2\gamma_2}{m}{\langle k_z^2\rangle}_h,\nonumber\\
E_l &\approx& \frac{\gamma_1+2\gamma_2}{m}{\langle k_z^2\rangle}_l,
\end{eqnarray}
where ${\langle k_z^2\rangle}_{h/l}$ is the average of $k_z^2$ with the $z$-component of the wave function.
With this approximation, 
the energy denominator in expressions of perturbation theory reads
\begin{equation}
E_l-E_h=
\frac{\gamma_1}{2m}\left({\langle k_z^2\rangle}_l-{\langle k_z^2\rangle}_h\right)+\frac{\gamma_2}{m}\left({\langle k_z^2\rangle}_l+{\langle k_z^2\rangle}_h\right).
\label{engdenomlh}
\end{equation}
It is important to remark here that this energy denominator does not depend on the quantum numbers of the inplane motion.
To simplify the notations above, we denoted all the quantum numbers by a single index, $h$ (or $l$).
Below, we shall single out the quantum number referring to the $z$-component of the wave function.
Thus, we shall replace $E_l-E_h$ in Eq.~(\ref{engdenomlh}) by $E_n^l-E_{n_0}^h$, where $n$ and $n_0$ 
denote quantum numbers of the motion along $z$ and the superscripts $h$ and $l$ indicate to which type of hole the expression belongs.
The $z$-components of wave functions associated with subband energies $E_n^l$ and $E_{n_0}^h$ are denoted by
$\psi_n^l(z)$ and $\psi_{n_0}^h(z)$, respectively.

We average over the motion along $z$ and obtain
\begin{eqnarray}
H_{\rm eff} &=& \langle H_{\it hh}\rangle+\langle\Delta H_{\it hh}\rangle,\nonumber\\
\langle\Delta H_{\it hh}\rangle &=& -\sum_{n}\left\langle \psi_{n_0}^h\right| H_{\it hl}\left|\psi_n^l\right\rangle 
\frac{1}{E_{n}^l-E_{n_0}^h}\left\langle \psi_n^l\right|H_{\it lh}\left|\psi_{n_0}^h\right\rangle,\nonumber\\
\end{eqnarray}
where $\langle\dots \rangle\equiv \left\langle \psi_{n_0}^h\right|\dots\left|\psi_{n_0}^h\right\rangle$ 
and $n_0$ denotes the number of the heavy-hole subband for which we derive the effective 2D Hamiltonian.
In practice, $n_0$ can be one of the several first subbands and in the main text we set $n_0=1$.
Further, the derivation continues as follows,
\begin{eqnarray}
\langle\Delta H_{hh}\rangle &=& -\sum_{n}
\frac{\left\langle \psi_{n_0}^h\right| k_z\left|\psi_n^l\right\rangle \left\langle \psi_n^l\right|k_z\left|\psi_{n_0}^h\right\rangle}{E_{n}^l-E_{n_0}^h}
\nonumber\\
&&\times
\frac{3\gamma_3^2}{m^2}(k_x\sigma_y+k_y\sigma_x)^2.
\end{eqnarray}
Next, note that
\begin{equation}
(k_x\sigma_y+k_y\sigma_x)^2=k_x^2+k_y^2-i\left[k_x,k_y\right]\sigma_z.
\end{equation}
Since $\left[k_x,k_y\right]=-i(\hbar e/c)B_z$, we obtain
\begin{equation}
\langle\Delta H_{hh}\rangle = -\frac{\gamma_{h,n_0}}{2m}
\left(k_x^2+k_y^2-\frac{e\hbar}{c}\sigma_zB_z \right),
\end{equation}
where
\begin{equation}
\gamma_{h,n_0} = \frac{6\gamma_3^2}{m}\sum_{n}
\frac{\left\langle \psi_{n_0}^h\right| k_z\left|\psi_n^l\right\rangle \left\langle \psi_n^l\right|k_z\left|\psi_{n_0}^h\right\rangle}{E_{n}^l-E_{n_0}^h}.
\label{gamheqdef}
\end{equation}
Therefore, we obtained a g-factor renormalization and a mass renormalization.
The heavy-hole g-factor $g_{z}$ changes to
\begin{equation}
g_{z,n_0}=-6\kappa-\frac{27}{2}q+2\gamma_{h,n_0}.
\end{equation}
The inplane mass reads
\begin{equation}
m_{\parallel,n_0}^{h} =\frac{m}{\gamma_1+\gamma_2-\gamma_{h,n_0}}.
\end{equation}


\begin{thebibliography}{99}
\bibitem{note1}
We choose the pseudospin basis as in Ref.~\onlinecite{spinselective};
namely,
$\left|\uparrow\right\rangle_{h}=\left|3/2,-3/2\right\rangle$
and
$\left|\downarrow\right\rangle_{h}=\left|3/2,+3/2\right\rangle$.
\end{thebibliography}
\end{document}